\documentstyle[aps,epsfig]{revtex}
\newcommand{\be}{\begin{equation}}
\newcommand{\ee}{\end{equation}}

\begin{document}
\draft
\title{Competition of Branches}
\author{M. B. Hastings}
\address{
CNLS, MS B258, Los Alamos National
Laboratory, Los Alamos, NM 87545, hastings@cnls.lanl.gov 
}
\date{May 16, 2001}
\maketitle
\begin{abstract}
We consider a general model of branch competition that automatically
leads to a critical branching configuration.  This model is
inspired by the $4-\eta$ expansion of the dielectric breakdown model (DBM),
but the mechanism of arriving at the critical point
may be of relevance to other branching systems as well, such as fractures.
The exact solution of this model clarifies the direct renormalization 
procedure used for the DBM, and demonstrates nonperturbatively the existence of 
additional irrelevant operators with complex scaling dimensions 
leading to discrete scale invariance.  The anomalous exponents
are shown to depend upon the details of branch interaction; we
contrast with the branched growth model (BGM) in which these
exponents are universal to lowest order in $1-\nu$, and show
that the BGM includes an inherent branch interaction different
from that found in the DBM.  We consider
stationary and non-stationary regimes, corresponding to 
different growth geometries in the dielectric-breakdown model.  
\vskip2mm
\end{abstract}
Diffusion-limited aggregation\cite{dla} produces complicated fractal
structures by branch competition.  
Recently, by considering the dielectric breakdown 
model\cite{dbm} (DBM),
a controlled renormalization group was developed within a $4-\eta$ expansion
for this class of models\cite{me}.  This expansion is based on considering
an aggregate as a collection of strictly one-dimensional branches; at
$\eta=4$, the aggregate consists of a single branch.
Given that the probability of a single branch pair surviving for time $t$
is proportional to $1/t^\nu$, to use the terminology of the branched
growth model\cite{bgm}, a direct renormalization group was developed
in an expansion in $1-\nu=(4-\eta)/2$.  It was found that, without fine 
tuning, the system arrives at a critical point characterized by a 
scale-invariant tip-splitting rate so that the renormalized
probability of a branch surviving for
a time $t$ is proportional to $1/t$ and the rate of branch production
and branch death balance at all scales.  However,  branching structures are 
common in other physical systems, and a similar $1/t$ kernel has been
found in fracture systems\cite{fracture}.

Thus, we will consider a more general model of branch competition,
inspired by the $4-\eta$ renormalization group.  This model will be
exactly solvable, clarifying the direct renormalization procedure employed
for the DBM.  
We will find that the exponents depend on the bare kernel and on 
the details of branch interaction, becoming trivial as the $\nu\rightarrow 1$.
Recently\cite{halsey}, it was shown that within the branched growth model (BGM)
the fractal dimension depends only on $\nu$ for small  $\nu$; here, this is 
shown not to be true for general models of branch interaction, though the 
fact that the numerical values of the exponents within the $4-\eta$\cite{me}
and branched growth\cite{halsey} expansions are similar is an indication
that the BGM is a useful approximation.  However,
we show that the physics of the BGM involves certain assumptions which
do not hold for the DBM, so that the BGM result is not exact for the
DBM.  The exact solution of the general model will also
reveal the presence of additional, irrelevant operators with complex
scaling dimension, indicating the presence of discrete scale
invariance, which has been argued to exist in DLA\cite{dsi}.

Finally, we will consider this model within two separate regimes.
One regime will be analogous to the steady state regime in the cylindrical 
geometry in the DBM, while the other will be a non-stationary state, which
corresponds to an initial condition of a single infinitely long vertical branch
in cylindrical geometry, over time scales much shorter than the time required 
for the aggregate to reach the scale of the cylinder.  We will find that
the exponents characterizing local fractal dimension and mass-radius
scaling are the same.  This indicates that the difference between these
dimensions observed in radial geometry\cite{mand}, as well as the non-trivial 
affine
exponents observed in the early stages of growth in cylindrical 
geometry\cite{ev} from
initial conditions consisting of a horizontal line, are due not just to
the non-stationary nature of the growth but also to the different
geometry and initial conditions.  Another motivation to consider different
geometries is to compare the branched growth model computation, performed
in a non-stationary regime, to the $4-\eta$ RG computation, performed
in a stationary regime.

{\it A Model of Branch Competition ---}
The dynamical state of the system at a given time $t$ will be
defined by a binary tree, with a set of times $t_i$, one for each branch
point in the tree, defining the time at which that branch pair was
produced.  There are two dynamical processes, tip-splitting which leads to the
production of additional branches and branch competition.
The first process is accounted for by assigning a rate $g$ at which
each tip splits, changing the topology of the tree by adding an additional
branch point at that tip with time $t_i=t$.  See Fig.~1.

The second process is accounted for by assigning a probability of
a branch point $i$ being removed from the tree due to competition of
branches.  When the branch point is created, one 
of the two branches below the branch point is randomly
designated as the weaker branch, and the other as the stronger.
When the branch point is removed, the weaker branch is removed from the tree.
See Fig.~2.  
We pick the probability of removing branch point $i$ to be
\be
\nu
\theta(t-t_i-1)\frac{1}{t-t_i},
\ee 
plus additional corrections due to branch interaction.  To include the effects 
of branch interaction, we pick a phenomenological model for the interaction.
We pick a scale factor $x<1$, and declare that if on the weaker branch 
there exists another branch point with branching time exactly equal 
to $t-x(t-t_i)$ then the
weaker branch will be removed.  In this case, the lower branch point
will have lived
for a time $x$ shorter than the branch point $i$.

This model has the essential features of the $4-\eta$ renormalization group.
One difference is that within that procedure the details of 
branch competition had to
be determined numerically by integrating the trajectories of
several competing branches, where
here they may be determined phenomenologically by the scale factor $x$.  
Another difference has to
do with the time scale.  Within the DBM, each tip has a given growth measure
and the tip-splitting rate and probability of removing branch points
are proportional to the growth measure at the tip normalized by the total
growth measure for the cluster.  Here, we instead take all tips to have
the same growth measure (so that all tips have the same probability of
splitting) and we do not normalize the tip-splitting rate.  The lack of
normalization simply changes the overall time scale; the choice of the
same growth measure for all tips is simply taken to make the model
more tractable analytically and does not alter the essential physics.

In addition, in the $4-\eta$ renormalization group, branch points may also
be removed as a result of interactions with parent branches of similar 
scale, rather than just as a result of daughter branches.  We will consider 
this possibility later.

{\it Survival Probability ---}
Let a branch pair be created at time $t=0$.  
Define $s(t)$ to be the probability that the branch pair survives until
time $t$, assuming that the branch pair is not destroyed by removal of
one of its parents.  We find
\be
\label{st}
\partial_t s(t)=-\frac{\nu \theta(t-1)s(t)}{t}
-g (1-x) s(t) s(x t),
\ee
as $g (1-x) s(x t) {\rm d}t$ is the probability of creating
a branch point at time $(1-x) t$ which survives until time $t$.
The growth rules have been chosen so that Eq.~(\ref{st}) is exact; the
process of creating and then removing a branch point does
not effect the distribution of branches on the remaining stronger branch below
that branch point.

Searching for a scaling solution, suppose 
$g s(t)=A/t$.
We find $1-\nu=(1-x) A/x$.  The constant $A$ sets the
probability, in time $t$, of producing a branch which remains in the tree
for time $t$; this is the dimensionless (scale invariant) 
tip-splitting rate\cite{me}.  As $\nu\rightarrow 1,A\rightarrow 0$.

This result can also be obtained by a direct renormalization procedure\cite{me}
in which one expands the survival probability in powers of $1-\nu$ and $g$:
to zeroth order in $g$, $g s(t)=A/t+A(1-\nu)\log{t}/t+{\cal O}(1-\nu)^2$.
Then, to first order in $g$ we find
$g s(t)=A/t+A(1-\nu)\log{t}/t-A^2 (1-x)\log{t}/(x t)$, so that a fixed
point of $t s$ is reached only for the given value of $A$.

To investigate the approach to the scaling solution, 
suppose instead
$g s(t)=\frac{A}{t} (1+f(t))$.
Linearizing Eq.~(\ref{st}) about $f=0$, we find
\be
\label{dcy}
\frac{\partial f(t)}{\partial \log{t}}=
-(1-\nu) f(x t).
\ee

Eq.~(\ref{dcy}) is translationally invariant in $\tau=\log{t}$,
and has solutions $f=e^{k \tau}$, where the eigenvalue
$k$ is the scaling dimension.  We find
\be
\label{scln}
k=-(1-\nu) x^{k}.
\ee

When $\nu\approx 1$, the eigenvalue with largest real part is
$k=-(1-\nu)+{\cal O}(1-\nu)^2$.  
Then, for $\nu \approx 1$, where this eigenvalue is small
we can make an approximation
that $f(x t)=f(t)$ and approximate the nonlinear problem by
$\partial_t s(t)=-\nu s(t)/t
-g (1-x) s^2(t)/x$,
which can be solved exactly for $s(t),t>1$ as 
$s(t)=(1-\nu)/(g(1-x) t/x + c t^\nu)$,
where $c$ is an arbitrary constant.

For $1-\nu$ small, Eq.~(\ref{scln}) has two solutions for real $k$, as well
as an infinity of solutions with complex $k$.  All these $k$ have
negative real part and describe irrelevant perturbations.  For $x^{1-\nu}=
e^{-1/e}$ the two real solutions merge, and for
$x^{1-\nu}<e^{-1/e}$ all solutions have complex $k$.
The presence of complex eigenvalues indicates that $s(t)$ has an oscillatory
behavior and that there is a discrete scale invariance in the
corrections to scaling.  The scale of this discrete scale invariance is
in general {\it not} equal to $x$, so that it is not simply
an artifact of the particularly simple form of branch interaction chosen,
but rather a result of the fact that branches separated by a finite range
of scales interact.
As $x^{1-\nu}$ is decreased, one finds a 
$k$ with vanishing real part when
$k=i \pi/(2\log{x})$, so $x^{1-\nu}=e^{-\pi/2}$.  For 
$x^{1-\nu}<e^{-\pi/2}$, there are
complex eigenvalues with positive real part which describe
{\it relevant} perturbations.  In this case, we have found numerically that
$f(t)$ is driven to a new fixed point with non-decaying log-periodic
oscillations, with scale not in general equal to $x$.
The model considered here is useful for analyzing these effects, which
are beyond perturbation theory in $1-\nu$.

Consider correlations: the probability, given that there is a branch
point at time $t_1$ remaining in the tree until time $t_1'$, that
there is another branch point at time $t_2$ remaining in the tree until 
time $t_2'$.  To study these, generalize the survival probability to
a function $s(t,t')$, the probability that a branch point created at
time $t$ survives until time $t+t'$.
We obtain the equation
\be
\partial_{t'} s(t,t')=-\frac{\nu \theta(t'-1) s(t,t')}{t'}
-g (1-x) s(t,t') s((1-x) t+x t',x t').
\ee
Defining $g s(t,t')=\frac{A}{t} (1+f(t,t'))$ and linearizing we find
\be
\partial_{t'} f(t,t')=
-\frac{(1-\nu)}{t'} f(t+(1-x) t',x t').
\ee
This equation is translationally invariant in $t$ and so we look for
solutions
$f(t,t')=e^{i l t} g(t')$.  We find
$\partial_{t'} g(t')=
-\frac{(1-\nu)}{t'} g(x t') e^{i l x t'}$.  For $l=0$, this
is the same as Eq.~(\ref{scln}).  Generally, for $t'<<l^{-1}$, we find the
same discrete scale invariant solution as for Eq.~(\ref{scln}) as above.  
For $t'>>l^{-1}$, the perturbations $f(t,t')$ decay more 
rapidly due to the oscillations of the exponential.
Thus, the two time scales $t,t'$ are related, and we find discrete scale
invariance in both.

{\it Different Geometries---}
We now consider the local ``fractal
dimension" of the cluster.  The length of a branch is defined to be
$t-t_i$, where $t_i$ is the time at which the pair to which that branch
belongs was created.  The largest such $t-t_i$ sets a length scale for
the cluster.  The mass of a branch is defined to be the length of
that branch, plus the sum, over all branch points $j$ which lie within
that branch at time $t$, of $t-t_j$, plus the sum, over all branch points 
which were added to that branch and later removed, of the elapsed time
between the addition and removal of the branch point.

Define $m(t)$ to be the average mass of a branch which is removed from
the tree after a time $t$ has elapsed since its creation (hence, this
is a weaker branch).  By averaging over the weights of sidebranches,
we obtain the exact recursion relation
\be
\label{mrec}
m(t)=t+g\int\limits_{0}^{t} {\rm d} t' \int\limits_{1}^{x t'} 
{\rm d}t'' m(t'') \bigl(-\partial_{t''}s(t'')\bigr).
\ee
Assuming a power law $m(t)=t^D$, one finds
$\alpha (\alpha-1)=A x^{\alpha-1}.$
For $\nu\approx 1$, we find
\be
\label{D}
D=1+(1-\nu)\frac{x}{1-x}+{\cal O}(1-\nu)^2.
\ee

Consider two different geometries.  If we start the tree with a single
branch and follow the dynamics above, this is analogous to starting the DBM with
a single branch as a seed configuration and letting the cluster grow.  In this
case, Eq.~(\ref{D}) provides a scaling of the mass with the time.  
In another geometry, analogous to the cylindrical geometry, we modify the 
dynamics to always remove a
weaker branch if $t-t_i>T$, for some $T$ setting a scale.  In this case,
after an initial non-stationary regime lasting for a time of order $T$, the
mass of the cluster increases linearly with $t$, with a rate proportional
to $T^{D-1}$: the mass of the largest branches, times the probability of
producing such a branch.  Thus, within this model the mass-radius scaling
and local fractal dimension are the same up to a trivial difference of unity.
We can generalize the model by including a possibility of removing a
branch due to the presence of a parent branch of comparable size.  
Then, the probability of the earliest branch point $i$ surviving till time
$t$ will scale as $(t-t_i)^{-a}t_i^{-b}$ where $a+b=1$.  One will again find 
that
the average time required to produce a branch point surviving for time $t$
is of order $t$ and the mass-radius scaling and local fractal dimension
will again be equivalent.

{\it Comparison to Branched Growth Model---}
It has been shown\cite{halsey} that there exists a $1-\nu$ expansion
for the BGM similar to the $4-\eta$ expansion for the DBM.  One elegant
feature of this expansion is that the lowest order fractal dimension
is obtained without considering interaction of branches, but simply from
the bare constant $\nu$.  This seems surprising, as we have found within
the model above, and within the $4-\eta$ expansion for the DBM, that
unless we include branch interaction, the tip-splitting rate
grows at large scales for $\nu<1$, and a scale invariant fixed point is
not reached.

The resolution of this is that the BGM is defined in a way which inherently
includes effects of branch competition.  In the BGM, a branch is assumed to have
a probability $1/m^{\nu}$ of surviving until it reaches a {\it mass} $m$, while
within the model above the probability is defined in terms of the probability to
survive for a {\it time} $t$.  Now, given that a scale invariant fixed point can
only be reached if a branch has probability $1/t$ of surviving for time $t$,
then we must have the relation that $1/m^{\nu}=1/t$, so that
$m=t^{1/\nu}$, and the fractal dimension is $1/\nu$.

More formally, $m(t)=t+A t \log{t}$ to order $(1-\nu)^0,A^1$.  We do
not have an exact Eq.~(\ref{mrec}) for the BGM, but this equation is
still correct to lowest order.  One may still define a survival probability
$s(t)$ for a branch within the BGM, and 
$\partial_{t} s(t)=-\nu \partial_{t} m(t) s(t)/m(t)$
so $s(t)=1/t+(1-\nu) \log{t}/t-A \log{t}/t$ to order 
$(1-\nu)^1,A^1$, and we find $A=1-\nu=D-1$.

However, this particular form of branch competition in the BGM is not
that found in the DBM, as found by numerically following the evolution of three
branches.  In some circumstances production of a daughter branch
can actually reduce the competition of the parent branch with its 
sister\cite{me}; it is only when summing over all configurations that the 
increase in branch competition is obtained.
Further, the competition of branches which is inherent in the
BGM involves only an increased competition of a branch pair due to daughter
branches.  However, within the $4-\eta$ expansion, it was necessary to consider
the interaction of branches with daughter and parent branches (which is
in fact the strongest interaction numerically) to obtain the
correct result.  Thus, both the present toy model and the BGM are approximations
to the physics near $\eta=4$, although the BGM dynamics serves also as a good
approximation for $\eta=1$\cite{bgm}.

{\it Conclusion---}
We have examined a simple, solvable model for branching, finding a fixed
point with a scale invariant tip splitting rate.  The direct RG for the
model is exact at lowest order.  The DBM has similar behavior
and a similar perturbative RG.
The exact solution enables us nonperturbatively to find
additional irrelevant operators leading to a discrete scale invariance.  
We contrast the behavior of
this model, and the DBM, with that of the BGM, for which a similar $1-\nu$
expansion is available.

{\it Acknowledgements---}
I thank T. C. Halsey for discussions.
This work was supported by DOE grant 
W-7405-ENG-36.  

\begin{figure}[!t]
\begin{center}
\leavevmode
\epsfig{figure=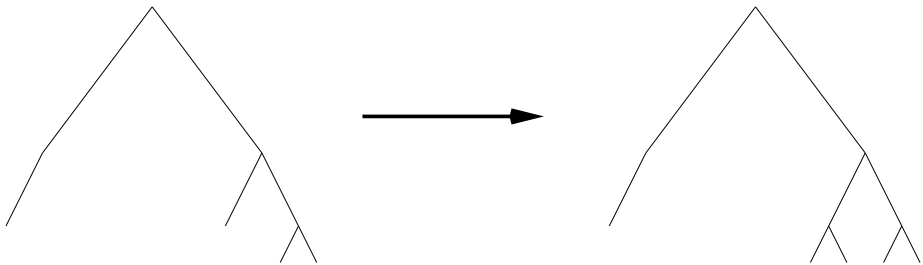,height=4cm,angle=0}
\end{center}
\caption{Creation of a new branch point.}
\end{figure}               
\begin{figure}[!t]
\begin{center}
\leavevmode
\epsfig{figure=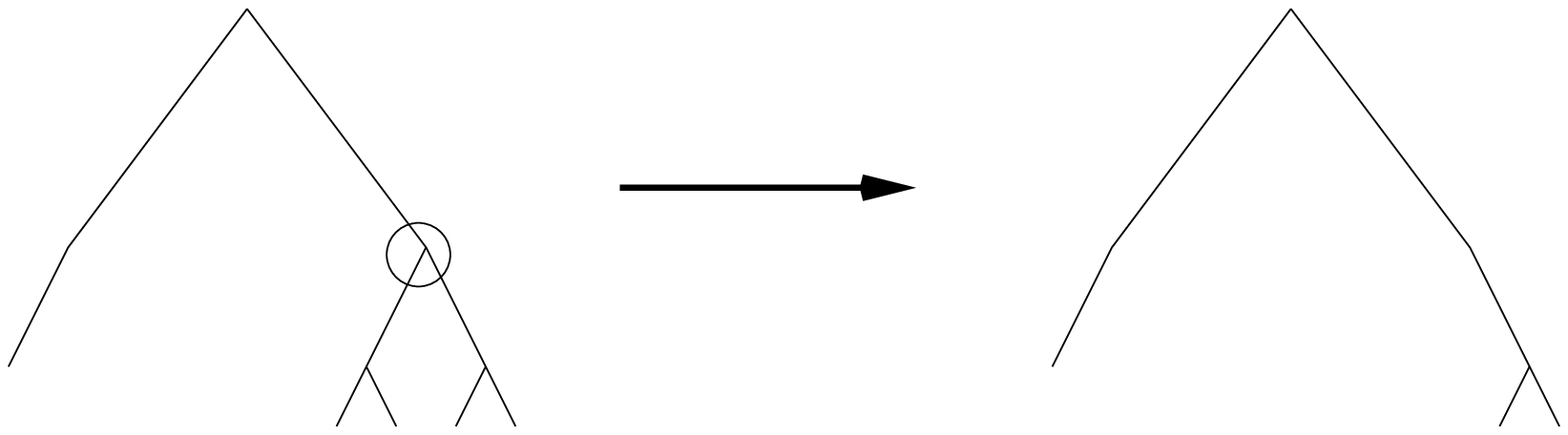,height=4cm,angle=0}
\end{center}
\caption{Removal of a branch below the circled branch point.}
\end{figure}               
\end{document}